# An affordable, wearable, fiber-free pulsed-mode diffuse speckle contrast flowmetry (PM-DSCF) sensor for noninvasive measurements of deep cerebral blood flow

**Chaebeom Yeo,**[a] **Xuhui Liu,**[a] **Mehrana Mohtasebi,**[a] **Faezeh Akbari,**[a] **Faraneh Fathi,**[a] **Guoqiang Yu**[a,*]

[a]Department of Biomedical Engineering, University of Kentucky, Lexington, Kentucky, USA

**Abstract**

**Significance:** Measuring cerebral blood flow (CBF) is crucial for diagnosing various cerebral diseases. An affordable, wearable, and fiber-free continuous-wave speckle contrast flowmetry (CW-DSCF) technique has been developed for continuous monitoring of CBF variations. However, its application in adult humans is limited by shallow tissue penetration.

**Aim:** To develop an innovative pulse-mode DSCF (PM-DSCF) system for continuous monitoring of CBF variations in adult humans.

**Approach:** The PM-DSCF utilizes an 808 nm laser diode and a small NanEye camera to capture diffuse laser speckle fluctuations caused by red blood cell movement in the brain (i.e., CBF). Operating in short-pulse mode (duty cycle < 5%), the system maximizes peak pulse light power for deeper tissue penetration, while ensuring that the average power density remains within ANSI safety standards for skin exposure. The PM-DSCF was evaluated on tissue-simulating phantoms and in adult humans.

**Results:** The maximum effective source-detector distance increased from 15 mm (CW-DSCF) to 35 mm (PM-DSCF). The PM-DSCF successfully detected CBF variations in adult brains during head-up-tilting experiments, consistent with physiological expectations.

**Conclusions:** Switching from CW mode to PM mode significantly increases the maximum tissue penetration depth from ~7.5 mm (CW-DSCF) to ~17.5 mm (PM-DSCF), enabling successful CBF measurements in adult humans.

**\*Corresponding Author**, Guoqiang Yu, E-mail: guoqiang.yu@uky.edu

# 1 Introduction

Continuous and longitudinal monitoring of deep tissue hemodynamics provides crucial information for diagnostic and therapeutic assessments of various diseases that manifest in large/deep tissue volumes such as tumors[1], burns/wounds[2], and injured brains[3]. An example of large/deep tissue volumes is the human brain, which lies beneath multiple tissue layers of scalp and skull. The average thickness of scalps and skulls in human adults is 10-15 mm[4, 5]. To effectively evaluate cerebral blood flow (CBF) beneath the top layers of scalp and skull, a minimum penetration depth of 10 mm from the surface of head is required. Moreover, integrating CBF measurements with wearable technologies allows for continuous and longitudinal monitoring of deep brain hemodynamics in conscious, freely behaving subjects, thus providing deep insights into cognitive processes and adaptive behaviors[6].

Various near infrared (NIR) diffuse optical techniques have been developed to measure CBF in human brains. A prominent example is the diffuse correlation spectroscopy (DCS)[7-15], which uses a point, continuous-wave (CW), coherent NIR light, and single-photon-counting avalanche photodiodes (APDs) to detect temporal fluctuations of diffuse laser speckles resulting from red blood cell motions in the brain (i.e., CBF). Conventional DCS systems employ source-detector (S-D) distances up to ~25 mm, resulting in a maximal penetration depth of ~12.5 mm (i.e., one-half of the S-D distance)[7, 15]. This penetration depth enables CBF measurements on cerebral cortex with large partial volume effects from the top-layered scalp and skull tissues. DCS performance has recently been enhanced through the use of longer wavelength illumination, interferometric detection, and time-resolved methods, achieving greater tissue penetration depths and higher sampling rates[16-21]. However, most DCS systems utilize large, expensive, discrete long-coherence lasers and APDs that are coupled to fiber-optic probes to detect CBF variations. As a result, DCS devices are generally large and expensive[8-10, 14, 15, 22-25] (>$50K for a 16-channel device[9, 10, 14, 15]). Moreover, the rigid fiber bundles used in DCS probes are sensitive to motion artifacts and constrain the subject's movement[16-21, 26].

Recent advancements have focused on using point, coherent NIR illumination for deep tissue penetration and cost-effective CCD/CMOS cameras to detect spatial fluctuations of diffuse laser speckles for CBF measurements. Based on the same principle, techniques such as diffuse speckle contrast analysis (DSCA), speckle contrast optical spectroscopy (SCOS), and our diffuse speckle contrast flowmetry (DSCF) replace expensive discrete APDs in DCS with 2D cameras, significantly improving sampling density, enhancing the SNR via spatial averaging across numerous pixels, and reducing device costs[6, 26-38]. More recently, pulsing methods have also been explored to enhance tissue penetration depths[29, 32]. However, most systems still rely on fiber optics to couple lasers and/or cameras to probes, making them susceptible to motion artifacts and limiting the subject's mobility. Furthermore, fiber coupling necessitates additional optical components, which complicates the instrumentation, increases its cost and size, and results in light loss and potential nonuniform detection.

To address these challenges, we have developed an innovative, low-cost, and fiber-free DSCF system (U.S. Patent #10/842,422, 2020, priority to July 21, 2017) with various sizes of wearable probes, enabling continuous and longitudinal measurements of CBF variations in animals and human neonates[6, 26, 30, 31, 34]. The CW-DSCF integrates low-cost, small laser diodes (e.g., $70, D780-30, US-Lasers) as focused point sources for deep tissue penetration and inexpensive, compact CMOS cameras (e.g., $150, NanEye 2D Black & White, Awaiba) as 2D detector arrays to measure CBF variations. A key innovation of CW-DSCF is its fiber-free design, with a wearable probe containing small laser diodes and compact cameras, connected to a portable electrical device via all flexible electrical wires. This unique design allows continuous monitoring in conscious subjects without fiber motion artifacts. Furthermore, using low-cost, fast, and compact 2D cameras for detection improves the sampling rate and SNR (via spatial sampling), enlarges head coverage, and reduces both the device's size and cost. Additionally, a dual-wavelength continuous-wave diffuse speckle contrast flow-oximetry (CW-DSCFO) system with time-multiplexed laser-diode illumination has been constructed for simultaneous measurements of CBF and cerebral oxygenation changes[26, 34]. The CW-DSCF/CW-DSCFO devices have been validated against conventional near-infrared spectroscopy and DCS on tissue-simulating phantoms, freely behaving rodents, neonatal piglets, and human preterm infants[6, 26, 30, 31, 34]. However, they cannot reliably detect deep tissue hemodynamics in adult human brains with thicker scalps and skulls, due to limited tissue penetration depth (~7.5 mm).

The goal of this study is to develop and validate an affordable, wearable, and fiber-free pulse-mode DSCF (PM-DSCF) system with deeper tissue penetrations for continuous monitoring of CBF variations in human adults. Briefly, the laser diode used in PM-DSCF operated in short pulse mode (duty cycle < 5%) instead of the CW mode used in our existing CW-DSCF. This change maximized peak pulse power for deep tissue penetration and high SNR, while limiting average power density to avoid thermal skin injury. The PM-DSCF system was optimized and characterized using tissue-simulating phantoms with known optical properties, and its performance was compared to our CW-DSCF system. The maximum effective S-D distance with valid signals increased significantly from 15 mm (CW-DSCF)[30, 34] to 35 mm (PM-DSCF). As a result, the tissue penetration depth increased from approximately 7.5 mm to 17.5 mm, enabling CBF measurements in adult humans. The optimized PM-DSCF system was subsequently used alongside a conventional DCS system to measure CBF changes concurrently and continuously in healthy adult volunteers during 70° head-up-tilting (HUT) experiments. Notably, both the PM-DSCF (S-D distance of 35 mm) and DCS (S-D distance of 25 mm) systems detected similar CBF variations in the adult brain during HUT experiments.

## 2 Materials and Methods

*2.1 PM-DSCF System*

*2.1.1 PM-DSCF Device*

Figure 1 shows the affordable, wearable, fiber-free PM-DSCF system designed based on our published CW-DSCF[30]. The PM-DSCF included a wearable probe consisting of a small laser diode as the point source and a tiny NanEye 2D camera as the 2D detection array (Fig. 1a). A small collimator (A230TM-B, effective focal length: 4.51 mm, numerical aperture: 0.55, outer diameter: 9.24 mm, working distance: 2.53 mm, anti-reflection coating: 650 - 1050 nm, Thorlabs) was placed in front of the small laser diode (L808P1000MM, Ø9 mm, 1 W, 808 nm, Thorlabs) to deliver a collimated light spot (diameter = 4.45 mm) on target tissue surface. A diffuser was placed between the collimator and the scalp to further ensure skin safety.

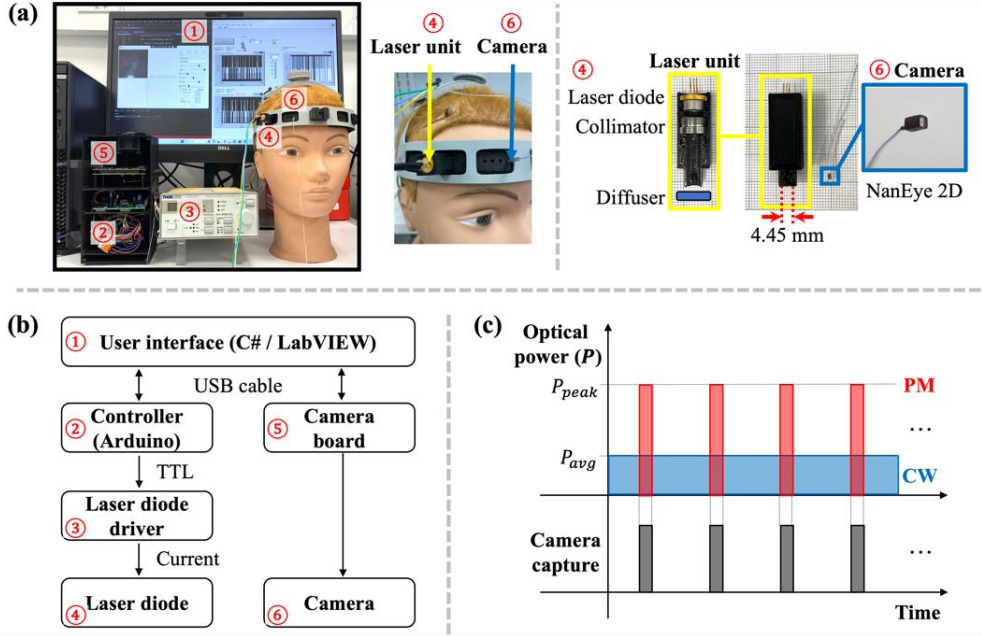

**Fig. 1:** An affordable, wearable, fiber-free PM-DSCF system for deep tissue blood flow measurements. **(a)** A headband of PM-DSCF placed on the subject's forehead. The PM-DSCF device included a user interface ①, an Arduino controller ②, a customized laser diode driver ③, a laser diode ④, a camera board (NanEye USB 2.2, Awaiba) ⑤, and a NanEye camera ⑥. **(b)** A schematic diagram of PM-DSCF device. **(c)** Synchronization of the pulsed laser and NanEye 2D camera in the PM-DSCF.

An ultra-small CMOS camera (NanEye 2D Black and White, Awaiba; Dimensions: $1 \times 1$ mm², Resolution: $250 \times 250$ pixels, Pixel Size: $3 \times 3$ μm², Bit Depth: 10 bits, Electronic Amplification (Gain): 2, Power Consumption: 4 mW) was utilized for detecting diffuse speckle contrasts. The NanEye camera drive includes a function that automatically scales the 10-bit data captured by the camera to 16-bit through interpolation. While this interpolation does not increase the dynamic range of detection, it enhances precision by leveraging the 16-bit format. A tiny lens with an effective focal length of 0.66 mm was integrated to the camera, which enabled an appropriate field of view (FOV) of $4 \times 4$ mm² at the working distance of 3.5 mm.

The laser diode probe and NanEye 2D camera were confined by a bendable foam pad and installed on the subject's forehead through a wearable headband. By adjusting the built-in dials on the headband, these source and detector components were firmly attached on subject's forehead.

The S-D distance between the laser diode and camera sensor could be changed to adjust tissue penetration depth.

The laser diode and camera were connected to a portable DSCF device via flexible electrical wires (fiber-free). As previously reported[6, 26, 30, 31, 34, 38], the DSCF operation was divided between a custom C# graphical user interface (GUI) and a LabVIEW™ (National Instruments) control panel installed on a computer (Fig. 1a). The C# GUI served as the primary interface, controlling the camera exposure and displaying real-time results (Fig. 1b). A LINX module and application logic in LabVIEW™ facilitated communication between the computer and an Arduino board to control the laser diode driver (LDC240C, ±4 A, Thorlabs), generating pulsed light with a preset intensity level and pulse width. Synchronization of the camera with the laser control module was achieved through transmission control protocol/internet protocol (TCP/IP) communication between C# and LabVIEW™ using the computer's loopback address (Fig. 1b). An optimal delay between the C# and LabVIEW™ controls was experimentally determined to align the camera exposure with the laser pulses. Additionally, any noticeable noises caused by misalignment were excluded from the data analysis. For experiments with varying camera exposure times, the laser pulse width was adjusted to ensure complete overlap, minimizing light waste (Fig. 1c).

*2.1.2 Pulse-mode Strategy*

The goal of this study is to increase tissue penetration depth of DSCF for continuous monitoring of CBF variations in human adults. Increasing S-D distance results in deeper tissue penetration depth but lower SNR due to greater light attenuation[30, 34]. The SNR refers to the ratio of the detected light intensity to the dark noise. To enhance measurement SNR, one can increase the laser light intensity or prolong the camera exposure time. However, high-intensity laser illumination can cause thermal injury to the skin. Conversely, increasing exposure time may degrade laser speckle contrast, leading to reduced measurement sensitivity for capturing up fast CBF changes[39, 40]. Measurement sensitivity refers to the ability of DSCF device to detect small changes in CBF.

We employed a pulse-mode strategy in the DSCF system that utilized high-intensity illumination for a short pulse duration while maintaining the average power density below the ANSI standard (ANSI Z136.1). According to this ANSI standard, the maximum power density at 808 nm should be less than 0.32 W/cm² on the tissue surface. Thus, the laser diode in the PM-DSCF operated in short pulse mode with a duty cycle of <5% and an average power density of ≤0.28 W/cm². Furthermore, the instantaneous peak power was consistently maintained at a constant level across all experiments with varying exposure times. Although the maximum sampling rate of NanEye camera is 50 Hz, the sampling rate of PM-DSCF was set to 1.65 Hz to ensure sufficient SNRs[30, 34] and maintain laser safety with pulse illumination.

*2.1.3 Data Processing*

In our CW-DSCF, laser speckle contrast was quantified in the spatial domain ($K_s$) within a pixel window of 7 × 7 pixels[6, 26, 30, 31, 34]. $K_s$ was calculated using the ratio of standard deviation ($\sigma$) to mean value ($\mu$) over light intensities ($I$) of 49 pixels (i.e., $K_s = \sigma(I)/\mu(I)$). A sliding window was applied over all pixels on the selected region-of-interest (ROI) to generate a 2D map of $K_s$. These $K_s$ values were then averaged to generate an averaged $K_s$ with improved SNR. A blood flow index ($BFI$) was subsequently defined as $(1/K_s^2)$[41, 42]. The relative changes in $BFI$ ($rBFI$ or rCBF) were determined by normalizing $BFI$ values to its baseline before physiological changes.

After examining the collected raw intensity images, we selected an ROI comprising 6,400 pixels (80 × 80) at the center of the NanEye camera sensor (250 × 250 pixels) for PM-DSCF data analysis, ensuring the exclusion of shadowing artifacts caused by the probe wall within the camera's FOV. Previous studies have employed a single-window approach to calculate a single $K_s$, utilizing various larger pixel window sizes, including 5,000[27], 2,500[43], 2,401[44], 1,000[45], and 441[46]. Based on our comparison results from human studies (data not shown), the single-window approach exhibited stronger rCBF responses to HUT compared to the sliding-window averaging approach. Furthermore, rCBF responses measured using the single window approach closely aligned with DCS measurements. Consequently, this study adopted a single window of 6,400 pixels for $K_s$ calculation.

*2.2 Measurement Protocols on Tissue-simulating Phantoms*

Figure 2a shows the experimental setup for evaluating the performance of PM-DSCF on tissue-simulating liquid phantoms. The use of tissue-simulating phantoms with known optical properties is a commonly accepted strategy for NIR instrument calibration and validation[41, 42]. The liquid phantom composed of Intralipid particles (Fresenius Kabi, Sweden), India ink, and water. India ink concentration in the phantoms regulates the absorption coefficient $\mu_a$ while Intralipid concentrations regulate the reduced scattering coefficient $\mu_s'$. Intralipid particles in the liquid phantom also provide Brownian motions (i.e., particle flow) to mimic movements of red blood cells in the brain[47, 48]. In this study, optical properties of liquid phantom were set to $\mu_a = 0.03$ cm$^{-1}$ and $\mu_s' = 9$ cm$^{-1}$. Note that the absorption coefficient $\mu_a$ of the phantom is lower than the typical absorption values for realistic tissue. This parameter will be optimized in future studies to improve accuracy and relevance.

For a fair comparison, the average power density of laser illumination on the phantom was set to ~0.28 Watt/cm$^2$ for both CW-DSCF and PM-DSCF measurements. The room light was dimmed during experiments to minimize ambient light impact. The CW-DSCF and PM-DSCF measurements were performed separately under each condition for 15 seconds at the sampling rates of 1.65 Hz. The mean value over 15 seconds was reported.

The first experiment was conducted on the tissue-simulating phantom under the room temperature of 23 ℃ (i.e., with a constant BFI) to compare the sensitivities, SNRs, and dark noises of CW-DSCF and PM-DSCF measurements with varied S-D distances and camera exposure times. The S-D distance varied from 10 mm to 40 mm with an interval of 5 mm by changing the camera location. At each S-D distance, camera exposure time varied from 1 ms to 15 ms with an interval of 5 ms.

The second experiment was performed on the same tissue-simulating phantom with varied temperature to compare the capabilities of CW-DSCF and PM-DSCF in detecting dynamic flow changes of Intralipid particles. The Brownian motion of Intralipid particles (i.e., particle flow) increases with the increase of phantom temperature[47, 48]. A heating plate (PC-620D, Corning) underneath the liquid tank was used to change the liquid temprature from 10 ℃ to 40 ℃ with an interval of 5 ℃. A thermometer sensor (Physitemp) was placed inside the liquid phantom to continuously monitor phantom temperature changes. To compare measurement sensitivities under different conditions, the S-D distance varied from 30 mm to 40 mm with an interval of 5 mm and the camera exposure time varied from 1 ms to 15 ms with an interval of 5 ms.

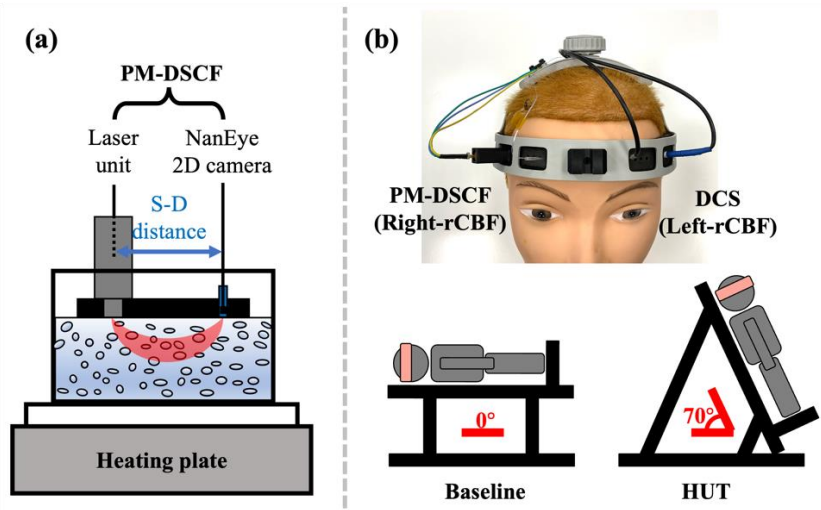

**Fig. 2:** Experimental setup for evaluating the performance of PM-DSCF in tissue-simulating phantoms and human adults. **(a)** A PM-DSCF probe was placed on the surface of Intralipid liquid phantom with known optical properties

to test and compare the measurement sensitivities and SNRs in detecting Intralipid particle flow. The heating plate underneath the liquid tank was used to manuplate the liquid temprature for changing Intralipid particle flow. **(b)** Top: A hybrid probe was placed on the the subject's forehead for concurrent PM-DSCF and DCS measurements of CBF changes in human adults. Bottom: CBF was continously monitored by the PM-DSCF and DCS devices during the 70° HUT experiment.

*2.3 Measurement Protocols in Human Adults*

Concurrent PM-DSCF and DCS measurements were performed in 6 healthy adult subjects (2 males and 4 females, age = 30.33 ± 2.88 years) to test the capability of wearable PM-DSCF in measuring deep CBF variations during 70° HUT experiments (Fig. 2b). The study protocol was approved by the University of Kentucky Institutional Review Board (IRB #44104). Room temperature was maintained at 23 ℃ and room light was dimmed during experiments. The participant was asked to lie on a titling table (Hausmann Inc., USA) in a supine position. The headband integrated with the PM-DSCF and DCS probes was placed on the subject's forehead for CBF measurements at right and left hemispheres, respectively. The focus of this study was to explore the use of larger S-D distances in PM-DSCF to probe deeper CBF in the adult brain. Consistent with previous studies[49], we physically tightened the headband (Fig. 1a and Fig. 2b) to ensure effective light coupling of the DSCF probe with the head while displacing scalp blood away from the probe. This approach minimizes the impact of scalp blood flow on the measurements.

DCS measurements were conducted with a customized instrument consisting of a long coherent 785 nm laser source (DL785-100, CrystaLaser) coupled with a multi-mode fiber (diameter = 200 μm) and an APD module (SPCM-AQ4C, PerkinElmer) coupled with a single-mode detector fiber (diameter = 5 μm). An autocorrelation board (www.correlator.com) took the output from the APD and calculated the autocorrelation functions for reconstructing BFI. Details about DCS instrumentation and data analysis can be found from previous publications[14, 15, 22, 24, 25].

A case study was first conducted in one subject (Subject #4) to characterize the performance of two devices with the camera exposure time of 10 ms (PM-DSCF) and varied S-D distances of 25 mm, 30 mm, and 35 mm for both DCS and PM-DSCF. Based on the case study results, the exposure time of 10 ms and S-D distance of 35 mm for PM-DSCF were chosen to measure CBF variations during HUT in 6 human subjects. However, the S-D distance of DCS was set to 25 mm due to its limited SNRs with longer S-D distances. CBF was continuously measured by the two devices at the sampling rate of 1.65 Hz (PM-DSCF) and 1 Hz (DCS) for 15 minutes during HUT with three phases: 0° baseline for 5 minutes, 70° titling for 5 minutes, and 0° recovery for 5 minutes. DCS data were interpolated to 1.65 Hz using a 1D-data interpolation function in MATLAB (Interp1) to align to the PM-DSCF data for comparisons. Regression correlation analyses were performed to evaluate correlations between the PM-DSCF and DCS measurements.

## 3    Results

*3.1 PM-DSCF Demonstrates Better Performance than CW-DSCF on Tissue-simulating Phantoms*

The stability of the laser diode was assessed by continuously monitoring the camera-detected intensity (digital counts) and the derived rBFI values in Intralipid liquid phantoms under controlled room temperature conditions for ten minutes. Results from three repeated measurements indicated coefficients of variation of $\leq 2.3\%$ and $\leq 10.2\%$ for camera intensity and $\leq 0.6\%$ and $\leq 7.6\%$ for derived rBFI at 5% and 10% duty cycles, respectively. These findings confirm that shorter duty cycles result in more stable outputs. Additionally, it was observed that the laser diode requires approximately five minutes of warm-up time prior to experiments to ensure stable power output.

Figure 3 presents a comparison of the CW-DSCF and PM-DSCF measurements in tissue-simulating phantoms under controlled room temperature conditions, focusing on sensitivities and SNRs as a function of the S-D distance and camera exposure time. As expected, the digital counts detected with the laser turned on were significantly higher than the dark counts recorded with the laser off (Fig. 3a and Fig. 3b). A closer examination of the dark noise zone reveals that the dark counts exhibit a linear dependence on exposure time (data not shown). The detected light intensities decreased with the increase of S-D distance and the decrease of camera exposure time for both CW-DSCF and PM-DSCF measurements. In comparison of CW-DSCF (Fig. 3a) and PM-DSCF (Fig. 3b), the intensity decay curves shifted from left (with shorter S-D distances) to right (with larger S-D distances) remarkably, indicating the improvement on tissue penetration depth.

As shown in Fig. 3c, SNRs decreased with the increase of S-D distance and the decrease of camera exposure time for both CW-DSCF and PM-DSCF measurements. The effective S-D distances with sufficient SNRs of $\geq 6$ dB and intensities below the camera saturation level (i.e., digital count $< 60,000$) were 15-25 mm (with camera exposure times of $\geq 5$ ms) for CW-DSCF measurements, and 20-35 mm (with camera exposure times of 5 ms and 10 ms) for PM-DSCF measurements.

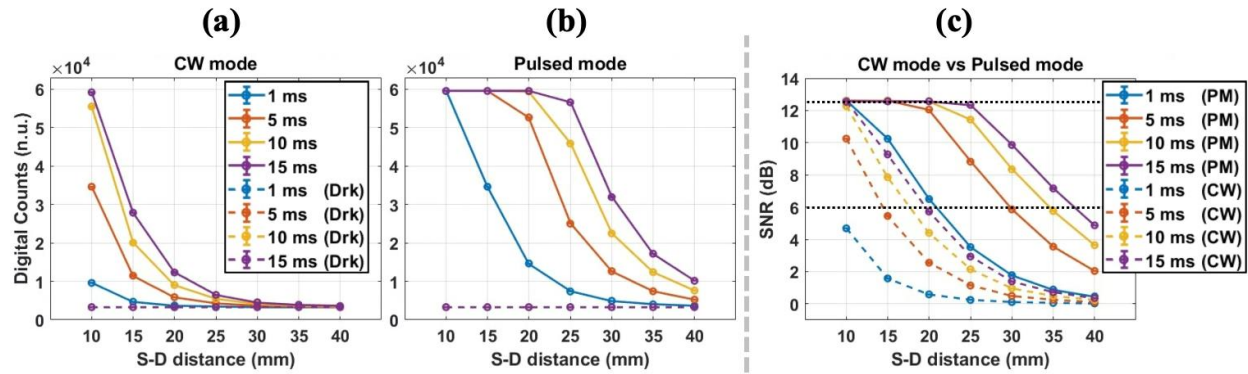

**Fig. 3:** Comparison of CW-DSCF and PM-DSCF measurement sensitivities and SNRs on tissue-simulating phantoms with varied S-D distances and camera exposure times. **(a)** Average light intensity (digital counts) and dark-noise counts detected by the CW-DSCF. **(b)** Average light intensity (digital counts) and dark-noise counts detected by the PM-DSCF. **(c)** Comparison of measurement SNRs achieved by the CW-DSCF and PM-DSCF. The two dashed

linesindicate the valid SNR range (6-12 dB), where the detected intensities remained well above the noise level (i.e., dark count) and below the camera saturation threshold (digital count < 60,000).

*3.2 PM-DSCF Captures Dynamic Flow Changes with Large Penetration Depths on Tissue Simulating Phantoms*

Figure 4 shows the results comparing CW-DSCF and PM-DSCF measurements of dynamic flow changes in tissue-simulating phantoms, induced by changing phantom temperature. As shown in the top panel (Fig. 4a-4c), CW-DSCF with S-D distances ≥ 30 mm lacked sufficient sensitivity to detect BFI changes. Sensitivity was observed with S-D distances ≤ 15 mm (data not shown), aligning with our published results[30, 34]. In contrast, PM-DSCF, with camera exposure times ≥ 5 ms and S-D distances of 30-40 mm, successfully detected BFI changes (Fig. 4d-4f), demonstrating deeper penetration capability compared to CW-DSCF. The extremely low $R^2$ values for bold cases in Table 1 indicate insensitivity of PM-DSCF for BFI detection at camera exposure times ≤ 5 ms and S-D distances ≥ 35 mm. Based on these findings, PM-DSCF measurements require camera exposure times ≥ 10 ms at the larger S-D distances (30 and 35 mm) to achieve sufficient SNR for BFI detection.

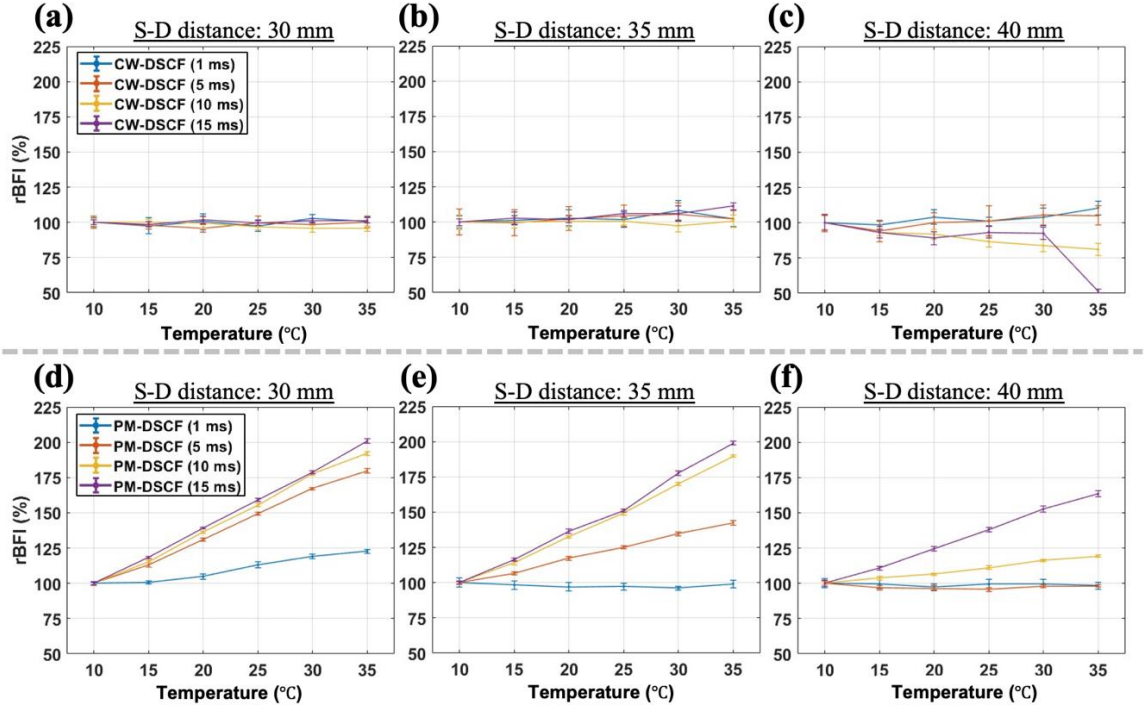

**Fig. 4:** CW-DSCF and PM-DSCF measurements of dynamic flow changes (rBFI) in tissue-simulating phantoms, induced by changing phantom temperature. **(a)**-**(c)** CW-DSCF measurements of rBFIs (means ± standard deviations) with S-D distances of 30, 35, and 40 mm, respectively. **(d)**-**(f)** PM-DSCF measurements of rBFIs (means ± standard deviations) with S-D distances of 30, 35, and 40 mm, respectively. The error bars represent the standard deviations of rBFI over a 7-second measurement period at each temperature, during which approximately 12 frames of data were collected.

**Table 1**: Linear regression correlations between the phantom temperatures and PM-DSCF measured rBFI values (Fig. 4d-4f).

| S-D distances | Exposure times | $R^2$ | P values | $X_1$ (Slope) | $X_0$ (Intercept) |
|---|---|---|---|---|---|
| 30 mm | 1 ms | 0.93 | $< 10^{-5}$ | 1.01 | 87.30 |
|  | 5 ms | 0.99 | $< 10^{-5}$ | 3.30 | 66.06 |
|  | 10 ms | 0.99 | $< 10^{-5}$ | 3.80 | 60.72 |
|  | 15 ms | 0.99 | $< 10^{-5}$ | 4.03 | 58.92 |
| 35 mm | **1 ms** | **0.04** | **0.20** | **-0.06** | **99.34** |
|  | 5 ms | 0.99 | $< 10^{-5}$ | 1.75 | 81.75 |
|  | 10 ms | 0.99 | $< 10^{-5}$ | 3.60 | 61.66 |
|  | 15 ms | 0.99 | $< 10^{-5}$ | 3.96 | 57.74 |
| 40 mm | **1 ms** | **0.01** | **0.54** | **-0.03** | **99.66** |
|  | **5 ms** | **0.03** | **0.21** | **-0.04** | **98.48** |
|  | 10 ms | 0.98 | $< 10^{-5}$ | 0.78 | 91.85 |
|  | 15 ms | 0.99 | $< 10^{-5}$ | 2.57 | 73.76 |

Note: Most p-values are less than $10^{-5}$, except for those bolded.

*3.3 PM-DSCF Enables Detection of rCBF Variations in Adult Human Brains*

Figure 5 shows the concurrent DCS and PM-DSCF measurement results of rCBF changes in left and right hemispheres of a illustrative subject during 70° HUT, using the S-D distances of 25, 30, and 35 mm, respectively. The camera exposure time was set to 10 ms for PM-DSCF measurements. The subject was given sufficient resting time (>1 hour) between the three measurements. PM-DSCF detected higher CBF responses to HUT as S-D distances increased. PM-DSCF measurements with different S-D distances exhibited more consistent right-rCBF responses to HUT with less noises, compared to the DCS measurements of left-rCBF. The DCS measurement with the short S-D distance of 25 mm detected similar rCBF changes like the PM-DSCF measurement with the same S-D distance (Fig. 5a). However, DCS measurements with the longer S-D distances of 30 mm and 35 mm exhibited remarkably decreased detection sensitivities and increased detection noises (Fig. 5b and Fig. 5c).

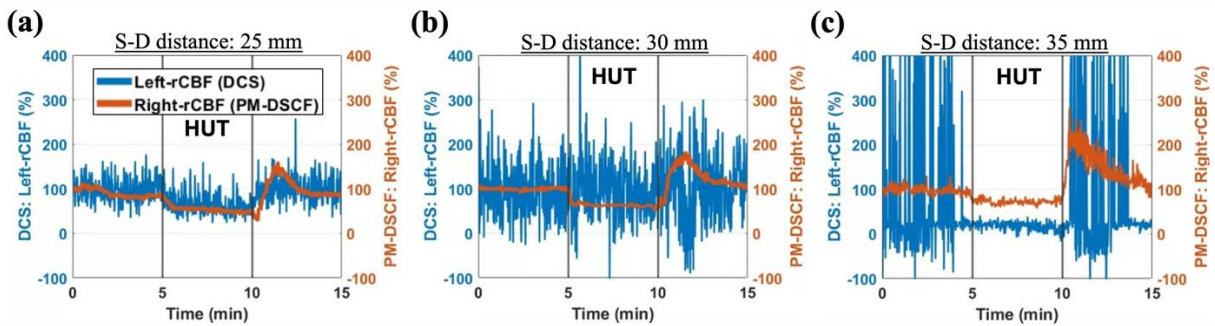

**Fig. 5:** Concurrent DCS and PM-DSCF measurements of rCBF changes in left and right hemispheres of a human adult (Subject #4) during 70° HUT, with S-D distances of **(a)** 25 mm, **(b)** 30 mm, and **(c)** 35 mm, respectively.

Figure 6 shows group average results of rCBF changes during 70° HUT from 6 subjects. The PM-DSCF measured right-rCBF with the S-D distance of 35 mm and exposure time of 10 ms while the DCS measured left-rCBF with the S-D distance of 25 mm. PM-DSCF measurements exhibited greater rCBF responses to HUT compared to DCS measurements (Fig. 6a). Analysis of the mean percent difference from the baseline (100%) during the HUT and recovery periods, based on PM-DSCF results, revealed values (means ± standard errors) of 45.23 ± 9.17% and 20.99 ± 35.20%, respectively. Similarly, analysis based on DCS results showed values of 44.13 ± 15.14% and 0.59 ± 23.86%, respectively. The linear regression correlation between the two measurements was significant ($R^2 = 0.75$, $p < 10^{-5}$, Fig. 6b). Furthermore, the regression slope was 1.25, indicating that PM-DSCF measurements have better sensitivity compared to DCS measurements.

The overshot in CBF observed during the recovery phase of the HUT experiment resembles post-occlusive reactive hyperemia, a phenomenon driven by tissue ischemia followed by reperfusion. During the HUT recovery phase, this reactive hyperemia restores normal perfusion to brain regions that experienced reduced flow during head-up tilting. The response is mediated by a transient reduction in sympathetic tone and activation of local vasodilatory mechanisms, leading to a temporary overshoot in CBF. Notably, our previous study using DCS also documented this CBF overshot[9]. The hyperemic responses detected by the DCS are less pronounced than those detected by the PM-DSCF (Fig. 6a). This difference is primarily due to the lower sensitivity of the DCS in deeper brain regions, especially when the tightened headband displaces scalp blood away from the probe.

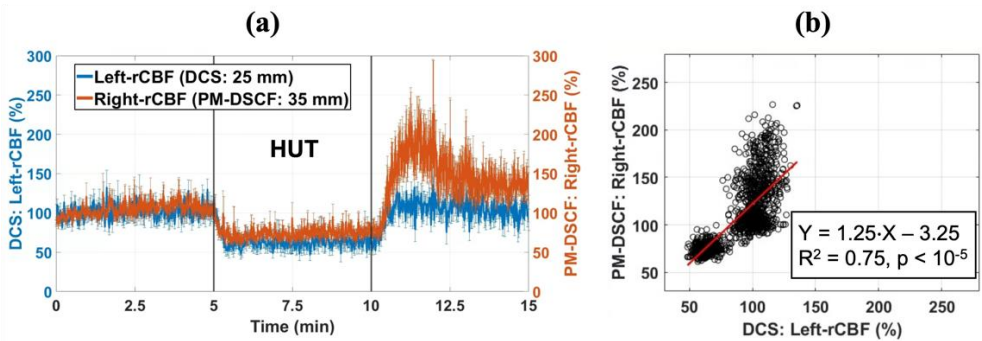

**Fig. 6:** Group average results from concurrent DCS and PM-DSCF measurements of rCBF changes in left and right hemispheres of 6 human adults during 70° HUT. **(a)** Time-course changes of left-rCBF (by DCS with the S-D distance of 25 mm) and right-rCBF (by PM-DSCF with the camera exposure time of 10 ms and S-D distance of 35 mm) during HUT. The error bars represent the standard errors over 6 subjects at each measurement time point. **(b)** Linear regression correlation between the averaged right-rCBF (PM-DSCF) and left-rCBF (DCS) from 6 subjects.

## 4   Discussion and Conclusions

Continuous monitoring of CBF holds promise for timely diagnosis and therapeutic management of many cerebral and neurovascular diseases. Portable near-infrared spectroscopy technologies have been utilized for continuous monitoring of CBF variations in relatively deep brains[7-13, 15, 26, 28, 29, 32, 37, 38, 50, 51]. However, most systems employ rigid fiber-optic probes for light delivering and detection, which constrain the subject's movement[7-15, 22, 28, 29, 32, 37, 38, 45, 50, 51]. Supported by the National Institute of Health, we have developed an innovative wearable fiber-free CW-DSCF probe with varied sizes for continuous monitoring of CBF variations in conscious animals and human neonates[6, 26, 30, 31, 34, 38]. Particularly, we have reported results from a comparison study with concurrent DCS and CW-DSCFO measurements in neonatal piglets, where motion artifacts were observed exclusively in the DCS measurements[26]. Despite this important progress, significant issues with the CW-DSCF remain, including low SNRs and limited tissue penetration depths (up to ~7.5 mm).

This paper reports an affordable, wearable, fiber-free PM-DSCF with improved SNR and tissue penetration depth for CBF measurements in human adults (Fig. 1). The PM-DSCF shares many similarities to the CW-DSCF, including wearable fiber-free sensor and low-cost portable device. Unlike CW-DSCF, PM-DSCF operates in short pulse mode (duty cycle < 5%) to maximize peak pulse light power while keeping the average power density below the ANSI standard (ANSI Z136.1) for skin safety. As a result, PM-DSCF enables measurements at larger S-D distances with sufficient SNR, allowing for the detection of CBF variations in adult humans with thicker scalp and skull layers.

We conducted a comprehensive comparison between CW-DSCF and PM-DSCF measurements in tissue-simulating phantoms (see Fig. 3, Fig. 4, and Table 1). The first experiment was conducted on an Intralipid phantom solution with a constant BFI to compare the sensitivities, SNRs, and dark noises of CW-DSCF and PM-DSCF measurements with varied S-D distances and camera exposure times (Fig. 3). The results in Fig. 3a are consistent with our previous publications, demonstrating that CW-DSCF achieves a maximum penetration depth of approximately 7.5 mm with an S-D distance of 15 mm[28,30,31]. In contrast, the results in Fig. 3b show that, with optimized exposure times, PM-DSCF achieves a maximum penetration depth of approximately 17.5 mm with an S-D distance of 35 mm.

The second experiment was performed on the same phantom with varying temperature to compare the capabilities of CW-DSCF and PM-DSCF in detecting dynamic flow changes of Intralipid particles. The results in Fig. 4a-4c demonstrate that CW-DSCF with large S-D distances of 30-40 mm cannot detect any changes in BFI. In contrast, the results in Fig. 4d-4f show that, with optimized exposure times, PM-DSCF with large S-D distances of 30-40 mm successfully detects dynamic BFI changes corresponding to the phantom temperature variations. Notably, PM-DSCF measurements with optimized exposure times at S-D distances of 30 mm and 35 mm were highly consistent (Fig. 4d and Fig. 4e). These results demonstrate the sensitivity of PM-DSCF in detecting dynamic flow changes in deep tissues (up to ~17.5 mm).

Concurrent PM-DSCF and DCS measurements were then performed in six healthy adult subjects to test and compare the capability of PM-DSCF in measuring deep CBF variations during 70° HUT experiments (Fig. 2b). PM-DSCF measurements with S-D distances of 25-35 mm exhibited more consistent rCBF responses to HUT and less noise compared to DCS measurements at a 25 mm S-D distance (Fig. 5a and Fig. 6a). In contrast, DCS measurements at larger S-D distances (30 mm and 35 mm) showed significantly reduced detection sensitivities and increased noise (Fig. 5a and Fig. 5c). These findings suggest that PM-DSCF offers better sensitivity and SNR than DCS. Additionally, no temperature changes on the scalp were noticed during the HTU experiment, and no complaints were reported by participants, indicating that both PM-DSCF and DCS measurements are safe and comfortable.

Some concerns and limitations of the present study have been identified. The statistics of speckle patterns depend on the spatial and temporal coherence of the light source, as well as the roughness of the surface or medium. A highly coherent laser source (such as a monochromatic laser) can generate speckle patterns with high speckle contrast, largely independent of the surface properties. These are referred to as normal speckle patterns. However, multimode lasers with partial coherence (like the one used in this study) can also generate speckle patterns, albeit with reduced speckle contrast and greater dependence on the surface or tissue structure[31, 52].

Pulse broadening in time-domain systems generally occurs during propagation through a diffuse medium, impacting the speckle contrast when the light's coherence length is insufficient. In our PM-DSCF approach, pulse durations range from 1 to 15 milliseconds, matching the camera exposure times (1-15 ms). These durations are considerably longer than the pulse durations typically used in time-domain systems, which usually range from picoseconds to nanoseconds. Similar millisecond-level durations have been used in NIRS systems, including continuous wave[49, 53-55] and frequency-domain[56-58] approaches, to enable fast sampling across multiple S-D channels. Recent advancements in DCS technologies, utilizing short illuminations of 10-100 ms, have enabled rapid measurements of pulsatile CBF associated with the cardiac cycle, albeit at the expense of reduced SNR[59-64]. In a separate study, we evaluated the impact of laser coherence length on CBF measurements using diffuse optical techniques[65]. Unlike DCS, DSCF does not require a laser with an excessively long coherence length. Overall, the effects of pulse broadening and coherence length shortening are not major concerns for our PM-DSCF approach with millisecond-level pulse durations.

In this study, we selected an ROI of 6400 pixels (80 × 80 pixels) at the center of the NanEye camera sensor (with a total resolution of 250 × 250 pixels) for data analysis, aiming to improve the SNR through spatial averaging across the pixels within the ROI. The limitation of utilizing only a portion of the total resolution can be mitigated by optimizing the PM-DSCF probe design to reduce shadowing artifacts caused by the probe wall. Additionally, the SNR can be further improved by using compact cameras with a higher pixel count, as the improvement from spatial averaging scales with the square root of the total number of pixels averaged.

As highlighted by recent publications[32, 66], low photon flux levels near the noise threshold are frequently encountered in human CBF measurements. Under these conditions, camera noise and nonidealities can significantly impact fiber-based SCOS measurements. The authors systematically evaluated various noise sources, including shot noise, read noise, and spatial inhomogeneity noise, and successfully mitigated these factors to enable accurate calculation of the BFI. However, the present study utilized the conventional method to analyze speckle contrasts (Section 2.1.3). Future studies will address this limitation by identifying various noise sources specific to the NanEye camera and, if necessary, implementing noise reduction methodologies to further enhance the accuracy of CBF measurements.

In summary, the results from the tissue-simulating phantoms and human adult heads (Fig. 3-6 and Table 1) demonstrate that switching from CW mode to PM mode with optimized exposure time significantly improves maximum effective S-D distances from 15 mm to 35 mm. Correspondingly, the maximum penetration depth of PM-DSCF reaches approximately 17.5 mm (half of 35 mm). This notable enhancement in tissue penetration depth facilitates CBF measurements in human adults. Future studies will focus on optimizing the system to develop and commercialize a low-cost, user-friendly, wireless, wearable, and fiber-free PM-DSCF system for continuous monitoring and instant reporting of cerebral hemodynamic variations in both comatose and conscious subjects, with the goal of real-time diagnosis and therapeutic monitoring of various cerebral vascular diseases and neurological disorders.

*Disclosures*

None.

*Code, Data, and Materials*

The data that support the reported findings are available from the corresponding author upon reasonable request.

*Acknowledgments*

We acknowledge partial financial support from the National Institutes of Health (NIH) #R56-NS117587, #R01-EB028792, #R01-HD101508, #R21-HD091118, #R21-NS114771, #R41-NS122722, #R42-MH135825 (G. Y.). The content is solely the responsibility of the authors and does not necessarily represent the official views of NIH or University of Kentucky.

**Caption List**

**Fig. 1:** An affordable, wearable, fiber-free PM-DSCF system for deep tissue blood flow measurements. **(a)** A headband of PM-DSCF placed on the subject's forehead. The PM-DSCF device included a user interface ①, an Arduino controller ②, a custimized laser diode driver ③, a laser diode ④, a camera board (NanEye USB 2.2, Awaiba) ⑤, and a NanEye camera ⑥. **(b)** A schematic diagram of PM-DSCF device. **(c)** Synchronization of the pulsed laser and NanEye 2D camera in the PM-DSCF.

**Fig. 2:** Experimental setup for evaluating the performance of PM-DSCF in tissue-simulating phantoms and human adults. **(a)** A PM-DSCF probe was placed on the surface of Intralipid liquid phantom with known optical properties to test and compare the measurement sensitivities and SNRs in detecting Intralipid particle flow. The heating plate underneath the liquid tank was used to manuplate the liquid temprature for changing Intralipid particle flow. **(b)** Top: A hybrid probe was placed on the the subject's forehead for concurrent PM-DSCF and DCS measurements of CBF changes in human adults. Bottom: CBF was continously monitored by the PM-DSCF and DCS devices during the 70° HUT experiment.

**Fig. 3:** Comparison of CW-DSCF and PM-DSCF measurement sensitivities and SNRs on tissue-simulating phantoms with varied S-D distances and camera exposure times. **(a)** Average light intensity (digital counts) and dark-noise counts detected by the CW-DSCF. **(b)** Average light intensity (digital counts) and dark-noise counts detected by the PM-DSCF. **(c)** Comparison of measurement SNRs achieved by the CW-DSCF and PM-DSCF. The two dashed lines indicate the valid SNR range (6-12 dB), where the detected intensities remained well above the noise level (i.e., dark count) and below the camera saturation threshold (digital count < 60,000).

**Fig. 4:** CW-DSCF and PM-DSCF measurements of dynamic flow changes (rBFI) in tissue-simulating phantoms, induced by changing phantom temperature. **(a)**-**(c)** CW-DSCF measurements of rBFIs (means ± standard deviations) with S-D distances of 30, 35, and 40 mm, respectively. **(d)**-**(f)** PM-DSCF measurements of rBFIs (means ± standard deviations) with S-D distances of 30, 35, and 40 mm, respectively. The error bars represent the standard deviations of rBFI over a 7-second measurement period at each temperature, during which approximately 12 frames of data were collected.

**Fig. 5:** Concurrent DCS and PM-DSCF measurements of rCBF changes in left and right hemispheres of a human adult (Subject #4) during 70° HUT, with S-D distances of **(a)** 25 mm, **(b)** 30 mm, and **(c)** 35 mm, respectively.

**Fig. 6:** Group average results from concurrent DCS and PM-DSCF measurements of rCBF changes in left and right hemispheres of 6 human adults during 70° HUT. **(a)** Time-course changes of left-rCBF (by DCS with the S-D distance of 25 mm) and right-rCBF (by PM-DSCF with the camera exposure time of 10 ms and S-D distance of 35 mm) during HUT. The error bars represent the standard errors over 6 subjects at each measurement time point. **(b)** Linear regression correlation between the averaged right-rCBF (PM-DSCF) and left-rCBF (DCS) from 6 subjects.

**Table 1**: Linear regression correlations between the phantom temperatures and PM-DSCF measured rBFI values (Fig. 4d-4f).